\documentclass[3p,times,twocolumn]{elsarticle}
 \biboptions{comma,sort&compress}
 
\usepackage{graphicx}
\usepackage{ecrc}

\volume{00}

\firstpage{1}

\journalname{Nuclear and Particle Physics Proceedings}

\runauth{}

\jid{nppp}

\jnltitlelogo{Nuclear and Particle Physics Proceedings}

\usepackage{amssymb}
\usepackage[figuresright]{rotating}

\begin{document}

\begin{frontmatter}

%%
%%%%%%%%%%%%%%%%%%%%%%%%%%%%%%%%%%%%%%%%%%%%%%%%%
\title{On the  gauge-invariant  operator 
$A^2_{\min}$ in Euclidean Yang-Mills theories}
 % \corref{cor0}}
 \cortext[cor0]{Talk given at 19th International Conference in Quantum Chromodynamics (QCD 16),  4 - 8 july 2016, Montpellier - FR}
 \author[label1]{M.~A.~L.~Capri}
\ead{caprimarcio@gmail.com}
\author[label1]{D.~Fiorentini} 
\ead{diegofiorentinia@gmail.com}
\author[label1]{M.~S.~Guimaraes}
\ead{ msguimaraes@uerj.br}
\author[label1]{B.~W.~Mintz}
\ead{bruno.mintz@uerj.br}    
\author[label1]{L.~F.~Palhares} 
\ead{leticia.palhares@uerj.br} 
\author[label1]{S.~P.~Sorella\fnref{fn1}}
   \fntext[fn1]{Speaker, Corresponding author.}
\ead{silvio.sorella@gmail.com} 
\address[label1]{UERJ -- Universidade do Estado do Rio de Janeiro, Instituto de F\'isica -- Departamento de F\'{\i}sica Te\'orica -- Rua S\~ao Francisco Xavier 524,
20550-013, Maracan\~a, Rio de Janeiro, Brasil}

\pagestyle{myheadings}
\markright{ }
\begin{abstract}
We review our recent work on the gauge-invariant non-local dimension-two operator 
$A^2_{\rm min}$,  whose minimization is defined along the gauge orbit. Albeit non-local, 
the operator $A^2_{\rm min}$ can be  cast in local form through the introduction of an auxiliary Stueckelberg field. 
The whole procedure results into a local action which turns out to be renormalizable to all orders. 
\end{abstract}
% \begin{document}
%\begin{keyword}  


%\end{keyword}

\end{frontmatter}
%%%%%%%%%%%%
%\vspace*{-1.5cm}
\section{Introduction}

The study of operators of dimension two in Yang-Mills theories has already a relatively long history, as confirmed by the 
considerable amount of results obtained through
theoretical and phenomenological studies as well as from lattice
simulations \cite
{Cornwall:1981zr,Greensite:1985vq,Stingl:1985hx,Lavelle:1988eg,Gubarev:2000nz,Gubarev:2000eu,
Verschelde:2001ia,Kondo:2001nq, Kondo:2001tm,Dudal:2003vv,
Browne:2003uv,Dudal:2003gu, Dudal:2003by,
Dudal:2004rx,Browne:2004mk, Gracey:2004bk,
Li:2004te,Boucaud:2001st,Boucaud:2002nc,
Boucaud:2005rm, RuizArriola:2004en,Suzuki:2004dw,Gubarev:2005it,Furui:2005bu,Boucaud:2005xn,
Chernodub:2005gz,Boucaud:2005xn,Boucaud:2008gn,Pene:2011kg,Boucaud:2010gr,Blossier:2010ky,Dudal:2010tf,
Boucaud:2011eh,Blossier:2011tf,Blossier:2013te}.\\\\
In particular, the dimension two gluon condensate $\left\langle A_{\mu }^{a}A_{\mu
}^{a}\right\rangle $ has been much investigated in the Landau gauge. According 
to \cite{Gubarev:2000nz}, this condensate enters the operator
product expansion (OPE) of the gluon propagator. A combined OPE
and lattice analysis has shown that this condensate can account for the $%
1/Q^{2}$ corrections which have been reported \cite
{Boucaud:2001st,Boucaud:2002nc,Boucaud:2005rm,RuizArriola:2004en,Furui:2005bu,
Boucaud:2005xn,Boucaud:2008gn,Pene:2011kg,Boucaud:2010gr,Blossier:2010ky,Boucaud:2011eh,
Blossier:2011tf,Blossier:2013te}
in the running of the coupling constant and in the gluon
correlation functions. \\\\
An effective potential for $\left\langle
A_{\mu }^{a}A_{\mu }^{a}\right\rangle $ in Landau gauge has been obtained and
evaluated in analytic form at two loops in \cite
{Verschelde:2001ia,Dudal:2003vv,Browne:2003uv,Browne:2004mk,Gracey:2004bk},
showing that a nonvanishing value of $\left\langle A_{\mu
}^{a}A_{\mu }^{a}\right\rangle $ is favoured as it lowers the
vacuum energy. As a consequence, a dynamical gluon mass is
generated. We point out that, in the Landau gauge, the operator
$A_{\mu }^{a}A_{\mu }^{a}$ turns out to be $BRST$-invariant on shell, a
property which has allowed for an all-orders proof of its
multiplicative renormalizability  \cite{Dudal:2002pq,Gracey:2002yt}. \\\\
Dimension-two condensates also play an important role within 
the context of the Gribov-Zwanziger approach to confinement 
\cite{Dudal:2005na,Dudal:2007cw,Dudal:2008sp,Dudal:2011gd,Vandersickel:2012tz}  as well as for the 
formation of a dynamical gluon mass within the framework of the Dyson-Schwinger equations in Landau 
gauge, as reported in \cite
{Cornwall:1981zr, Aguilar:2008xm,Aguilar:2015bud}. These non-perturbative effects give rise to the 
so called decoupling solution for the gluon propagator 
\cite{Cornwall:1981zr,Dudal:2005na,Dudal:2007cw,Dudal:2008sp,Aguilar:2008xm,Fischer:2008uz}, 
{\it i.e.} to a propagator which exhibits positivity violation, while attaining a finite non-vanishing 
value at zero momentum. Until now, this behaviour is in very good agreement with the most recent lattice 
numerical simulations  \cite{Cucchieri:2007md,Cucchieri:2007rg,Cucchieri:2011ig,Oliveira:2012eh}. The  
generalization of these results to the linear covariant gauges has been worked out recently and can be found in  
\cite{Sobreiro:2005vn,Aguilar:2015nqa,Huber:2015ria,Capri:2015pja,Capri:2015ixa,Capri:2015nzw,Capri:2016aqq,Cucchieri:2009kk,Cucchieri:2011aa,Bicudo:2015rma}. \\\\Even if a large amount  of results has been obtained, many aspects related to dimension-two operators require further investigation as, for instance,  the issue of the gauge invariance,  a topic of pivotal importance in order to give a precise physical meaning to the corresponding 
condensates. This is precisely the aspect which has been addressed recently in  \cite{Fiorentini:2016rwx} and which will be reviewed in the present contribution.

\section{The gauge invariant operator $A^2_{min}$}

A genuine gauge-invariant dimension-two operator $A_{\min }^{2}$  can be constructed by minimizing  the functional
$\mathrm{Tr}\int d^{4}x\,A_{\mu }^{u}A_{\mu }^{u}$ along the gauge
orbit of $A_{\mu }$, see  \cite{Fiorentini:2016rwx} and refs. therein, namely
\begin{eqnarray}
A_{\min }^{2} &\equiv   &\min_{\{u\}}\mathrm{Tr}{\;}   \frac{1}{2} \int d^{4}x\,A_{\mu
}^{u}A_{\mu }^{u}\;,
\nonumber \\
A_{\mu }^{u} &=&u^{\dagger }A_{\mu }u+\frac{i}{g}u^{\dagger }\partial _{\mu
}u\;.  \label{Aminn0}
\end{eqnarray}
In particular, the stationary condition of the functional (\ref{Aminn0})  gives rise to a non-local transverse 
field configuration $A^h_\mu$, $\partial_\mu A^h_\mu=0$, which can be expressed as an infinite series in 
the gauge field $A_\mu$ \cite{Fiorentini:2016rwx}, {\it i.e.}
\begin{eqnarray}
A_{\mu }^{h} &=& {\cal P}_{\mu \nu} \left( A_{\nu }-ig\left[ \frac{1}{\partial ^{2}}\partial A,A_{\nu
}\right]  \right. \nonumber \\ &+&  \left. \frac{ig}{2}\left[ \frac{1}{\partial ^{2}}\partial A,\partial
_{\nu }\frac{1}{\partial ^{2}}\partial A\right]  \right)  +O(A^{3})      \label{ah}
\end{eqnarray}
where ${\cal P}_{\mu \nu}= (\delta_{\mu\nu} - \frac{\partial_\mu \partial_\nu}{\partial^2})$ is the transverse projector.   
The configuration $A_{\mu }^{h}$ turns out to be left invariant
by infinitesimal gauge transformations order by order in the gauge
coupling $g$ \cite{Lavelle:1995ty} 
\begin{equation}
\delta A_{\mu }^{h} =0\;,  \qquad 
\delta A_{\mu } =-\partial _{\mu }\omega +ig\left[ A_{\mu },\omega \right]
\;.  \label{gio}
\end{equation}
The gauge-invariant nature of expression (\ref{Aminn0})  can be made manifest by rewriting it in terms of the  field
strength $F_{\mu \nu }$. In fact, it turns out that \cite{Zwanziger:1990tn}
\begin{eqnarray}
A_{\min }^{2} = \frac{1}{2} \mathrm{Tr} \int d^4x A^h_\mu A^h_\mu =  -\frac{1}{2}\mathrm{Tr}\int d^{4}x\left( F_{\mu
\nu }\frac{1}{D^{2}}F_{\mu \nu }  \right. \nonumber \\
 + 2i\frac{1}{D^{2}}F_{\lambda \mu
}\left[ \frac{1}{D^{2}}D_{\kappa }F_{\kappa \lambda
},\frac{1}{D^{2}}D_{\nu }F_{\nu \mu }\right]  
\nonumber \\
-2i\left. \frac{1}{D^{2}}F_{\lambda \mu }\left[ \frac{1}{D^{2}}D_{\kappa
}F_{\kappa \nu },\frac{1}{D^{2}}D_{\nu }F_{\lambda \mu }\right] \right)
+O(F^{4})    \nonumber 
\end{eqnarray}
where the operator $({D^{2}})^{-1}$ 
denotes the inverse of the Laplacian $D^2=D_\mu D_\mu$ with $D_\mu$ being the 
covariant derivative \cite{Zwanziger:1990tn}. Let us also notice that, in the particular case of the Landau gauge 
$\partial_\mu A_\mu=0$, the gauge invariant quantity  $(A^h_\mu A^h_\mu)$ reduces to the operator $A^{2}$
\begin{equation}
(A^{h,a}_\mu A^{h,a}_\mu) \Big|_{\rm Landau} =  A^a_\mu A^a_\mu \;.      \label{land}
\end{equation}

\section{Construction of a local Lagrangian for $A^2_{min}$}

In order to construct a local action,  we start 
with the standard Faddeev-Popov action of Yang-Mills theory quantized in 
linear covariant gauges with the inclusion of the non-local gauge invariant  operator $(A^h_\mu A^h_\mu)$ as well as of a constraint enforcing the 
transversality of the field configuration $A_\mu^h$, {\it i.e.} we consider the action 
\begin{equation}
S = S_{FP}+
\int d^4x \left(\tau^{a}\,\partial_{\mu}A^{h,a}_{\mu}
+\frac{m^{2}}{2}\,A^{h,a}_{\mu}A^{h,a}_{\mu}\right)  \label{sact}
\end{equation} 
where $S_{FP}$ stands for the Faddeev-Popov action in linear covariant gauges   
\[
S_{FP} = \int d^{4}x\,\bigg(\frac{1}{4}\,F^{a}_{\mu\nu}F^{a}_{\mu\nu}+\frac{\alpha}{2}\,b^{a}b^{a}
+ib^{a}\,\partial_{\mu}A^{a}_{\mu}
+\bar{c}^{a}\,\partial_{\mu}D^{ab}_{\mu}c^{b}\bigg)    \nonumber
\] 
and where we have introduced the operator $(A^h_\mu A^h_\mu)$ through the mass parameter $m^2$. Also, the transversality of $A_\mu^h$ is enforced by the 
Lagrange multiplier $\tau^{a}$. \\\\The action (\ref{sact}) is still non-local, since the expression for $(A^h_\mu A^h_\mu)$  is an infinite sum of nonlocal
terms.  Nevertheless, expression (\ref{sact}) can be cast in local form  \cite{Fiorentini:2016rwx}  by means of 
the introduction of an auxiliary localizing Stueckelberg field $\xi^a$, whose role is to give, for each gauge 
field $A_\mu$, its corresponding configuration that minimizes the functional $A^2$, {\it i.e.}, 
$A^h_\mu$. This is most naturally implemented by defining a field $h$
which effectively acts on $A_\mu$ as a gauge transformation would act, in order to provide the 
minimizing configuration $A^h$, that is,
\begin{equation}\label{local_Ah}
A^{h}_{\mu} \equiv A^{h,a}_{\mu}\,T^{a}=h^{\dagger}A_{\mu}h+\frac{i}{g}h^{\dagger}\partial_{\mu}h.
\end{equation}
with
\begin{equation}
h=e^{ig\xi}=e^{ig\xi^{a}T^{a}},     \label{hxi}
\end{equation}
where $\{T^a\}$ are the generators of the gauge group $SU(N)$ and $\xi^{a}$ is a Stueckelberg field. 
\\\\Thus, by substituting the expression (\ref{local_Ah}) for $A^h$ in the action (\ref{sact}), 
we now have a local theory in terms of the field $\xi$. The price 
one has to pay to have such a local theory is a non-polynomial action. Indeed, by expanding 
(\ref{local_Ah}), one finds an infinite series whose first terms are
\begin{equation}\label{Ah_expansion}
 (A^{h})^{a}_{\mu}=A^{a}_{\mu}-D^{ab}_{\mu}\xi^{b}-\frac{g}{2}f^{abc}\xi^{b}D^{cd}_{\mu}\xi^{d}
 +\mathcal{O}(\xi^{3})\,,   
 \end{equation}
where 
\begin{equation}
 D_\mu^{ab} = \delta^{ab}\partial_\mu - gf^{abc}A_\mu^c 
\end{equation}
is the covariant derivative in the adjoint representation. 
\\\\The nonlocal expression (\ref{ah}) for 
$A^h_\mu$ in terms of the gauge field $A_\mu$  can 
be recovered by imposing the transversality condition $\partial_\mu A^h_\mu=0$, {\it i.e.} after taking 
the divergence of both sides of (\ref{Ah_expansion}), equating it to zero and solving for the Stueckelberg 
field $\xi^a$ \cite{Capri:2015ixa,Capri:2015nzw,Fiorentini:2016rwx}. Due to the transversality condition enforced by the Lagrange multiplier $\tau^a$,  the Stueckelberg field 
$\xi^a$ acquires now a specific meaning:  it is precisely the field which brings a generic gauge 
configuration $A_\mu$ into the gauge-invariant and transverse field configuration $A^h_\mu$ 
which minimizes the functional $A^2_{min}$. As shown in \cite{Fiorentini:2016rwx},  this feature,  
encoded in the term $\int d^4x\;\tau^{a}\,\partial_{\mu}A^{h,a}_{\mu}$, gives rise to deep differences 
between our construction and the standard Stueckelberg mass term, which is known to be a 
non-renormalizable theory which has to be treated as an effective field theory  \cite{Ferrari:2004pd}. 
\\\\An important property  
of $A^h_\mu$, as defined by eq.(\ref{local_Ah}), is its gauge invariance, that is,
\begin{equation}
A^{h}_{\mu} \rightarrow A^{h}_{\mu} \;,
\end{equation}
as can be seen from the gauge transformations  with $SU(N)$ matrix $V$
\[
A_\mu \rightarrow V^{\dagger} A_\mu V + \frac{i}{g} V^{\dagger} \partial_ \mu V
\;, \qquad h \rightarrow V^{\dagger} h    
\]
The local version of the action (\ref{sact}), in terms of the Stueckelberg field 
$\xi^a$, is thus given by
\begin{eqnarray}
S &=& S_{FP}+
\int d^4x \left(\tau^{a}\,\partial_{\mu}A^{h,a}_{\mu}
+\frac{m^{2}}{2}\,A^{h,a}_{\mu}A^{h,a}_{\mu}\right)\nonumber\\
&=& S_{FP}+\int d^4x \left[ \tau^a\left(A^{a}_{\mu}-D^{ab}_{\mu}\xi^{b}\right)\right]
 \nonumber\\
&&+ \;\;\frac{m^2}{2}\int d^4x \left(A^{a}_{\mu}-D^{ab}_{\mu}\xi^{b}\right)
\left(A^{a}_{\mu}-D^{ae}_{\mu}\xi^{e}\right)\nonumber\\
&& + \;\;\;\cdots
\label{S_local}
\end{eqnarray}
Due to the use of the auxiliary Stueckelberg field $\xi^a$, expression (\ref{S_local}) exhibits a non-polynomial character. 

\subsection{BRST invariance}

The local action $S$ enjoys an exact BRST symmetry \cite{Fiorentini:2016rwx}:
\begin{equation}
s S = 0 \;, \label{brst_s}
\end{equation}
where the nilpotent BRST transformations are given by 
\begin{eqnarray}
sA^{a}_{\mu}&=&-D^{ab}_{\mu}c^{b}\,,\nonumber \\
sc^{a}&=&\frac{g}{2}f^{abc}c^{b}c^{c}\,, \nonumber \\
s\bar{c}^{a}&=&ib^{a}\,,\nonumber \\
sb^{a}&=&0\,, \nonumber \\
s \tau^a & = & 0\,, \nonumber \\
s^2 &=0 &\;.    \label{brst}
\end{eqnarray}
From \cite{Dragon:1996tk}, for the Stueckelberg field we have,  with $i,j$ indices associated with a generic representation,
\begin{equation}
s h^{ij} = -ig c^a (T^a)^{ik} h^{kj}  \;, \qquad s (A^h)^a_\mu = 0  \;,  \label{brstst}
\end{equation}
from which the BRST transformation of the field $\xi^a$ can be evaluated iteratively, yielding
\[
s \xi^a=  - c^a + \frac{g}{2} f^{abc}c^b \xi^c - \frac{g^2}{12} f^{amr} f^{mpq} c^p \xi^q \xi^r + O(\xi^3)  
\] 
As shown in \cite{Fiorentini:2016rwx}, the BRST invariance of the action $S$, eq.(\ref{sact}),  can be translated into 
functional identities which can be used to show that $S$ is in fact renornalizable to all orders of 
perturbation theory.

\section{Conclusion} 

A gauge invariant dimension two operator can be introduced by minimizing  the functional
$\mathrm{Tr}\int d^{4}x\,A_{\mu }^{u}A_{\mu }^{u}$ along the gauge orbit, {\it i.e.}
\begin{equation}
A_{\min }^{2} =\frac{1}{2} \mathrm{Tr}\int d^{4}x\,A_{\mu }^{h}A_{\mu }^{h}\;,   \label{cc1}
\end{equation}
with $A_\mu^h$ the transverse configuration, $\partial_\mu A^h_\mu=0$, given in expression (\ref{ah}). \\\\Despite the highly non-local character, a fully local set up for both operators 
$(A^h_\mu A^h_\mu)$ and $A^h_\mu$ can be constructed, giving rise to a local and BRST-invariant action $S$, eq.(\ref{sact}), which turns out to be renormalizable to all orders of perturbation theory \cite{Fiorentini:2016rwx}.  Let us conclude by mentioning that, owing to the gauge invariance of $(A^h_\mu A^h_\mu)$ and $A^h_\mu$, the corresponding anomalous dimensions, $(\gamma_{(A^{h})^2}, \gamma_{A^{h}})$, turn out to be independent from the gauge parameter $\alpha$ entering the gauge fixing condition, 
 being given by \cite{Fiorentini:2016rwx}
\begin{eqnarray}
\gamma_{(A^{h})^2} &= &  \gamma_{A^{2}}\Big|_{\rm Landau}  =-\left(\frac{\beta(a)}{a}+\gamma^{\mathrm{Landau}}_{A}(a)\right)    \nonumber \\
\gamma_{A^{h}}& = & \gamma_{A^{h}}\Big|_{\alpha=0}=\gamma^{\mathrm{Landau}}_{A}(a)  \nonumber \\
 a&=& \frac{g^{2}}{16\pi^{2}}\ \label{ad} 
\end{eqnarray} 
where $(\beta(a), \gamma^{\rm Landau}_A(a))$ denote, respectively, the $\beta$-function and 
the anomalous dimension of the gauge field $A_\mu$ in the Landau gauge, corresponding to set the gauge parameter $\alpha$ to zero, $\alpha=0$.    One sees therefore that $(\gamma_{(A^{h})^2}, \gamma_{A^{h}})$ are not independent parameters of the theory.

\end{document}